\documentclass[12pt,aps,pre,onecolumn,showpacs,showkeys]{revtex4}
\usepackage{amssymb}
\usepackage{amsmath}
\usepackage{graphics}
\usepackage{amsfonts}
\usepackage{graphicx}
\usepackage{color}
\usepackage{soul}
\begin{document}

\title[Limiting periodic orbits with exponential and power law dynamics]{Constructive approach to limiting periodic orbits with exponential and power law dynamics}
\author{A. Provata }
\email{a.provata@inn.demokritos.gr}
\address{Institute of Nanoscience and Nanotechnology, National Center for Scientific Research ``Demokritos'',
GR-15310 Athens, Greece}

%
\begin{abstract}
In dynamical systems limit cycles arise as a result of a Hopf bifurcation, after a control parameter
has crossed its critical value. In this study 
we present a constructive method to produce dissipative dynamics which lead to stable periodic orbits
as time grows, with predesigned transient dynamics.
Depending on the construction method a) the limiting orbit can be a regular circle, an
ellipse or a more complex closed orbit and b) the approach to the limiting orbit can follow an exponential law or
a power law.
This technique allows to design nonlinear models of dynamical systems 
with desired (exponential or power law) relaxation properties.
\end{abstract}
 \date{Received: \today / Revised version: date}

\maketitle
\section{Introduction}
\label{sec:intro}

\par Limit cycles are on the bases of dissipative systems: they constitute 2D limiting 
orbits in the phase space and they are approached by the system trajectories as time goes to infinity. 
Normally, they arise after the system's control parameter passes through a Hopf bifurcation, 
also known as Poincar\'e--Andronov--Hopf bifurcation\cite{guckenheimer:1983,nicolis:1995,anishchenko:2002}. 
Limit cycles are frequently met in reaction diffusion systems in the modelling of oscillatory reactions\cite{nicolis:1977,prigogine:1967,lefever:1967,oregonator:1972}, 
in neuron dynamics in the modelling of the neuron spiking potentials 
\cite{fitzhugh:1961,nagumo:1962}, 
in ecology in the sustainable cycles of  predator-prey dynamics \cite{murray:2002, shabunin:2002}, 
in laser physics \cite{bohm:2015,kotwal:2017,shena:2017} and in general synchronization
phenomena \cite{bohm:2015,shena:2017,kuramoto:2002,abrams:2004,omelchenko:2015,schoell:2016}.

\par According to the theory of supercritical Hopf bifurcation, when the control parameter takes values below
 the bifurcation point the trajectories are attracted by the fixed point. 
On the bifurcation point the fixed point looses it stability, 
and above the bifurcation point the trajectories diverge 
away from the fixed point and get attracted by the emerging limit cycle \cite{guckenheimer:1983,nicolis:1995}. 
Although the picture of the Hopf bifurcation mechanism is well understood the nonlinearity of the equations 
rarely allows to have the direct solutions of the dynamical systems which produce the limiting orbit. 
In most cases only numerical integration of the differential equations is possible and this way 
we do not have control over the temporal approach to the limiting orbits or the relaxation properties of the trajectories. 

\par From the applications point of view it is always useful to have the exact time dependent 
equations which 
solve a dynamical system and define its temporal evolution. But, it is of equal importance to 
have the differential equations which produce the dynamics. Based on the differential equations
which describe the evolution of a single element, 
one can proceed to construct coupled systems, such as coupled neuron oscillators or coupled oscillatory reactions
\cite{anishchenko:2002,schoell:2016,tinsley:2012}. Networks of oscillators are
on the basis of many aspects of current research directed to biological, medical electronic, or physical applications.

\par It is useful to stress here the difference between the limit cycle dynamics 
and the conservative center dynamics \cite{nicolis:1995}. 
In the latter case it is possible to construct time independent
quantities which serves as the ``constants of motion'' and depend on the initial conditions. 
That is why these systems are termed as ``conservative'' and the trajectories, when unperturbed, will 
stay perpetually around the orbit dictated by the initial conditions. The former case, limit cycle dynamics, 
belongs in the general category of dissipative dynamics. This means that there is no conserved quantity and independently of the initial conditions, the system will tend to a specific limiting periodic orbit which is dictated by the dynamics
(and not by the initial conditions). In the current study we will concentrate on this dissipative case 
and will construct 2D dynamical systems which are analytically solvable and have predefined temporal approach to 
limiting periodic orbits (cycles).

\par As stated earlier, in most of the cases where a limit cycle arises after the system has 
crossed the critical point of the Hopf bifurcation the analytical
form of the trajectory is not known. This is partly due to the complex scheme of the limiting orbit and partly due to the complex 
temporal dependence as the trajectory curls around the limiting curve. For this reason we decided to create limiting orbits 
constructively, starting from the simplest cases and working on to the more complex schemes. 
This way the analytical solutions of the 
simplest cases naturally arise, while the more complex cases can be understood either 
as generalizations of the simple periodic orbits or as superpositions of several simple ones.

\par In the next section, Sec.~\ref{sec:exponential}, 
we present the construction technique using as a paradigmatic example
the case where the limiting set is a simple circle in 2D 
and the approach to the limiting trajectory follows an exponential law. 
In the same section we construct the dynamical systems 
leading to more complex closed limiting trajectories in 2D, 
such as an elliptic orbit.
In Sec.~\ref{sec:powerlaw} we construct dynamical systems 
which approach their limiting trajectory following a temporal power law.
This approach is slow as compared to the exponential case and only 
very late in time the shape of the limiting orbit emerges.
A summary of the results and open questions are provided in the concluding section.

\section{Exponential approach to limiting orbits in 2D}
\label{sec:exponential}
In this section, we present a constructive approach to circular and elliptic limiting
orbits, when the transient dynamics are of exponential type. To keep the construction pedagogical,
we start with the pure harmonic motion and compexify gradually the orbit radius in time,
Sec.~\ref{sec:exp-circle} and in shape, Sec.~\ref{sec:exp-ellipse}.

\subsection{Exponential approach: The circle as a limiting orbit}

\label{sec:exp-circle}
Consider a vector $\vec{R(t)}$ of constant magnitude $R$, whose center is fixed and its
endpoint moves around the circle with constant
angular velocity $\omega$. Without loss of generality, the origin is set at $(0,0)$ and
the equations describing a position of the end point $\left[ x(t),y(t)\right]$ of the vector $\vec{R(t)}$ with time are defined as:
\begin{subequations}
\begin{align}
x(t) &= R \cos (\omega t)\\
y(t) &= R \sin (\omega t)\\
x^2(t)& + y^2(t)=R^2.
\end{align}
\label{eqno01}
\end{subequations}
 The dynamical system which admits as solutions the above equations is the classical harmonic oscillator:
 \begin{subequations}
\begin{align}
\frac{dx}{dt}&=-  \omega y(t) \\
\frac{dy}{dt}&= \omega x(t) 
\end{align}
\label{eqno02}
\end{subequations}
This system is conservative and Hamiltonian. The energy (constant of motion) is conserved and only depends on the 
value of $R$. 
The variables $x(t)$ and $y(t)$ can be
considered as canonical coordinates (position, conjugate momentum), while the dynamics corresponds to a center. The system, depending 
on its initial conditions $R$, makes conservative cycles around the center $(0,0)$ with constant radius $R$.
\par Now consider the case, where the radius $r(t)$
 of the circle is not constant but starts at $R_0$ at $t=0$ and its value increases exponentially
to the value $R>R_0$, as time $t\to \infty$, in the following way: 
\begin{equation}
r(t)=R(1-Ae^{-at}).
\label{eqno03}
\end{equation}
In Eq.~\ref{eqno03}, $A$ is a positive constant, $0<A<1$ and $a$ is the exponential rate of the radius increase. 
Note that $A$ cannot take values greater than 1, because this leads to negative values for the radius $r(t)$, which is
unnatural.
At $t=0$ the radius is $R_0=R(1-A)$, while for $t\to \infty$ the radius tends exponentially to $R(t\to \infty)= R =$ const.
In this case we can envisage a trajectory starting with radius $R_0$ which expands continuously to $R$, keeping constant angular velocity
$\omega$. Note that in this case the angular velocity corresponds to the angle which was covered per unit time and 
not to the arc covered, because the
circle radius increases in time. Using Eq.~\ref{eqno03} the motion of the endpoint of the vector $\vec{R}(t)$ takes the form:
\begin{subequations}
\begin{align}
x(t) &= r(t) \cos (\omega t)\\
y(t) &= r(t) \sin (\omega t)\\
x^2(t)& + y^2(t)=r^2(t)=R^2\left( 1-Ae^{-at}\right) ^2.
\end{align}
\label{eqno04}
\end{subequations} 
\par To proceed to the dynamical system leading to these equations we note that the $x$ and $r$ temporal derivatives are connected
through the relation:
$dx/dt=\left( dr/dt \right)\cos (\omega t)-r(t)\> \omega \> sin(\omega t) $ and similarly for the $y-$variable. The derivative of
the changing radius $r(t)$ can be calculated from Eq.~\ref{eqno03}. Taking these into account the dynamical system which admits 
Eqs.~\ref{eqno04} as solution is:
\begin{subequations}
\begin{align}
\frac{dx}{dt} &= -ax+\frac{aR}{\sqrt{x^2+y^2}}x-\omega y\\
\frac{dy}{dt} &= -ay+\frac{aR}{\sqrt{x^2+y^2}}y+ \omega x
\end{align}
\label{eqno05}
\end{subequations}
Note that the exponential (temporal) approach to the circular orbit has changed
the original linear dynamics, Eqs.~\ref{eqno02}, to nonlinear, Eqs.~\ref{eqno05}. 
In the case that the radius is time independent Eq.~\ref{eqno01}c holds and 
consequently Eqs.~\ref{eqno04} reduce to Eqs.~\ref{eqno02}, the harmonic oscillator.

\par
\begin{figure}[ht!]
\begin{center}
\includegraphics[width=0.7\linewidth]{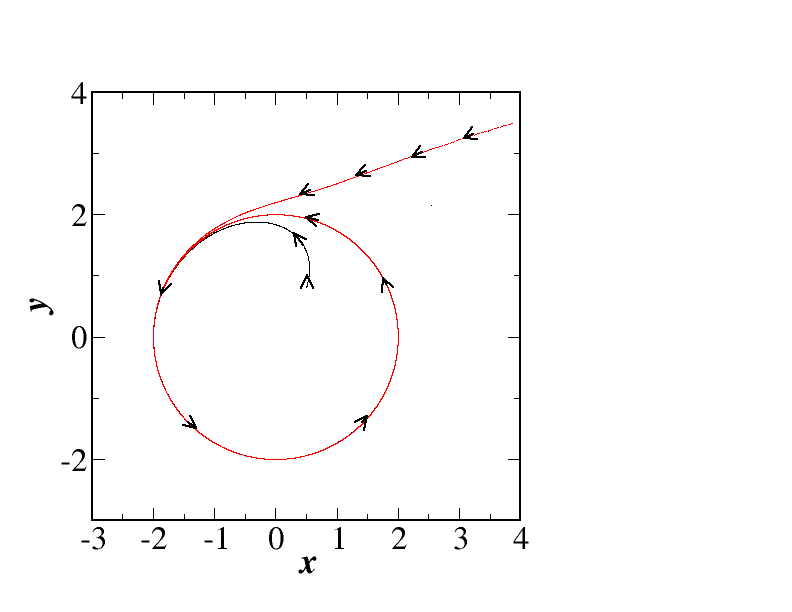}
\end{center}
\caption{(Color online) Phase space diagram representing the variables $x(t)$ and $y(t)$
as they approach their limit cycle, from numerical integration of Eqs.~\ref{eqno05}.
Parameters are $a=1$, $R=2$ and $\omega =0.3$. The black-solid line and the red-dashed line
correspond to different initial conditions.}
\label{fig:01}
\end{figure}

 Figure \ref{fig:01} depicts the phase space produced when Eqs.~\ref{eqno05} are numerically
integrated, with parameters $a=1$, $R=2$ and $\omega =0.3$, starting from different
initial conditions. Indeed, the resulting
phase space corresponds to the solutions Eq.~\ref{eqno04}, i.e., a circle with radius $R=2$
around the origin. 

We have then constructed a periodic motion in 2D with radius varying exponentially with rate $a$,
and tending asymptotically to a constant value as $t\to \infty$.
The scheme ~\ref{eqno05} is useful because it provides the nonlinear evolution dynamics of the $x(t)$ and $y(t)$
variables for which the analytical solutions are also known,  Eqs.~\ref{eqno04}. [In most cases involving limiting orbits 
it is rare to know both the analytical solutions and the evolution dynamics, at the same time.]
The variable $A$ is a control variable which defines the initial radius of the system. 
Namely, for $t=0$ the system is at $r(0)=R(1-A)$;
when $A=1$ the system starts from the center of the motion $r(0)=0$ and tends to $r(t\to \infty)=R$,
while for $A=0$ the system starts from $r(0)=R$ and remains there for ever (conservative motion). The variable $A$ 
is then a measure of the ``amount of dissipation'' in the system and 
quantifies the deviation of the motion from conservative dynamics. 
\par The parameter $a$ takes values in the interval $[0,\infty ]$ and is a control parameter 
which relates to the speed at which the trajectory approaches the limit cycle. When $a=0$ the trajectory
is on the limit cycle with constant radius $R(1-A)$. 
For small values of $a<<1$ the trajectory approaches fast the limiting curve,
while when $a>>1$ the trajectory takes a long time before approaching its limiting orbit. 

\par An equivalent way to write the system ~\ref{eqno05} is by defining the complex variable $z(t)=x(t)+iy(t)$.
System ~\ref{eqno05} reduces to
\begin{equation}
\frac{dz}{dt} = -az+\frac{aR}{|z|}z+i\omega z.
\label{eqno06}
\end{equation}

\par In an analogous way we can treat the case where the radius decreases 
from $R_0=R(1+A)>R$ to $R$ when $t\to\infty$, while its temporal dependence is:
\begin{equation}
r(t)=R(1+Ae^{-at}).
\label{eqno07}
\end{equation}

\par Next,  we complexify the limiting
orbit going beyond the simple circle but keeping the approach to the limiting orbit exponential.

\subsection{Exponential approach: Generalizing to elliptic limiting orbits}
\label{sec:exp-ellipse}

In this section the approach to the limiting orbit will be kept to exponential law,
but the final orbit will be elliptic, using two parameters, rather than 
the simple circular one introduced in the previous section.
 Without loss of generality, we consider the case where the $x$ and $y$ variables have different limiting values $R_1$
and $R_2$ as $t \to \infty$.  The temporal approach to the limiting values remains exponential with the same 
exponent $a$ for both variables and the origin is set again at $(0,0)$. 
The equations describing the position of the end point $\left[ x(t),y(t)\right] $ 
of the vector $\vec{R(t)}$ with time take the form:
\begin{subequations}
\begin{align}
x(t) &= R_1(1-Ae^{-at}) \cos (\omega t)\\
y(t) &= R_2 (1-Ae^{-at})\sin (\omega t)\\
\frac{x^2(t)}{{R_1} ^2}& + \frac{y^2(t)}{{R_2} ^2}=(1-Ae^{-at})^2.
\end{align}
\label{eqno11}
\end{subequations}
By taking the derivatives with respect to time of the above equations we find the dynamical system which 
admits as solutions these equations, namely:

\begin{subequations}
\begin{align}
\frac{dx}{dt} &= -ax+\frac{a}{\sqrt{x^2/R_1^2+y^2/R_2^2}}x-\omega \frac{R_1}{R_2} y\\
\frac{dy}{dt} &= -ay+\frac{a}{\sqrt{x^2/R_1^2+y^2/R_2^2}}y+ \omega \frac{R_2}{R_1} x
\end{align}
\label{eqno12}
\end{subequations}
The rate of dissipation only depends on $a$ and is, thus, equal in the $x-$ and $y-$ directions.

\par
\begin{figure}[ht!]
\centering
\includegraphics[width=0.7\linewidth]{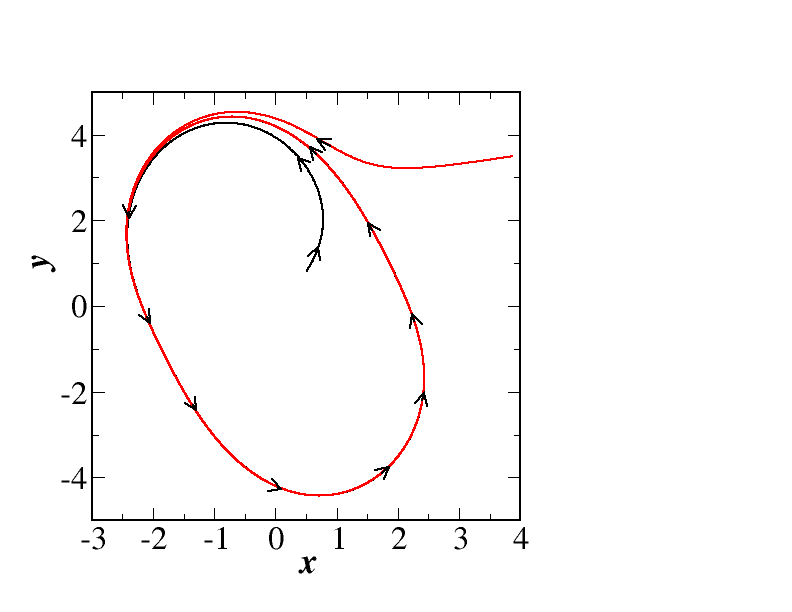}
\caption{(Color online) Phase space diagram representing the variables $x(t)$ and $y(t)$
as they approach their limiting elliptic orbit (numerical integration of Eqs.~\ref{eqno12}).
Parameters are $a=1$, $R_1=2$, $R_2=5$ and $\omega =0.3$. The black-solid line and the red-dashed line
correspond to different initial conditions.}
\label{fig:02}
\end{figure}

 In an analogous way to Fig.~\ref{fig:01}, Fig.~\ref{fig:02} depicts the phase space produced when Eqs.~\ref{eqno12} are numerically
integrated, with parameters $a=1$, $R_1=2$, $R_2=5$ and $\omega =0.3$, starting from different
initial conditions. The resulting
phase space corresponds to the solutions Eq.~\ref{eqno12}, i.e., a ellipse
around the origin. 

\par
\begin{figure}[ht!]
\centering
\includegraphics[width=0.7\linewidth]{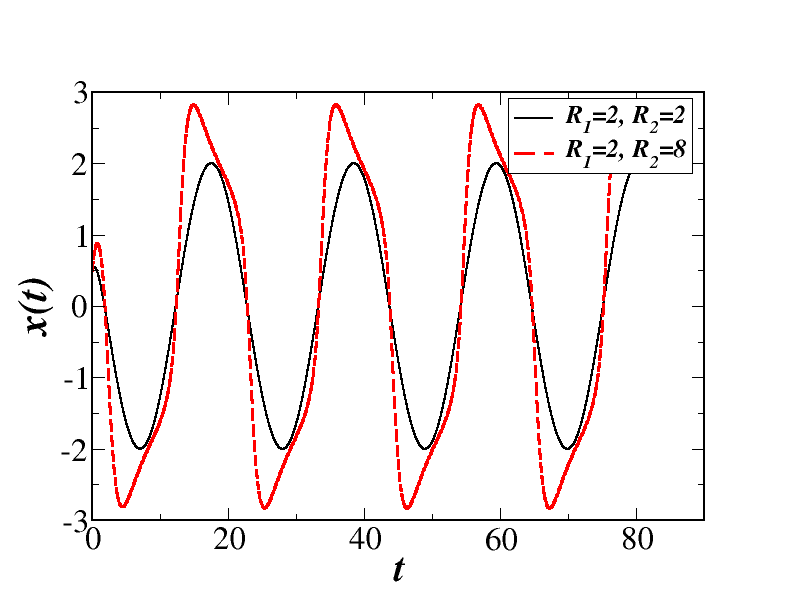}
\caption{(Color online) Temporal evolution of the variable $x(t)$ for different values
of $R_1$ and $R_2$, as indicated in the diagram, with common $\omega =0.3$ and common initial conditions.}
\label{fig:03}
\end{figure}

 The difference between the  semi-major and semi-minor axes of the ellipse, $R_1$ and $R_2$,
 gives rise to different shapes 
of the orbit, ranging from purely harmonic oscillations to spiking dynamics, 
most useful in neuron dynamics \cite{fitzhugh:1961,nagumo:1962}. In Fig.~\ref{fig:03} we
present the temporal evolution of $x(t)$ for harmonic orbits corresponding to $R_1=R_2$ (black, solid curve) 
and for elliptic orbits, $R_2>R_1$, (red, dashed line). Spiking dynamics is exhibited for large difference
in the two axes of the ellipse (see red, dashed curve). Although the two curves have identical periods/frequencies
the red one makes larger escapes in the $y-$direction, forcing the orbit to cover 
longer distances and to produce abrupt returns to the base.
Spiking dynamics is frequently observed in natural process, such as in heartbeat
dynamics, brain activity and other biological processes. In particular, the recently observed phenomenon 
of chimera states is usually manifested when the participating oscillators have spiking dynamics.
Chimera states \cite{kuramoto:2002,abrams:2004,omelchenko:2015,schoell:2016} are stable configurations in networks
consisting  of identical and identically coupled oscillators which spontaneously 
split into coexisting coherent and incoherent domains. These states are clearly manifested when the 
constituent oscillators are spiking limit cycles. Using the proposed approach we can now construct models
with predetermined spiking dynamics after an exponential relaxation
and we can study the formation of chimera states as a function of 
the ratio $R_1/R_2$ or $R_1-R_2$, the difference of the ellipse's axes.

It is possible to produce complex phase space if the two variables of the system have different frequencies
and different exponential approach to the steady state. In the general case it is not possible to write down
dynamical systems, equivalent to Eqs.~\ref{eqno05} and ~\ref{eqno12}, although one may write the explicit
form of the solutions, e.g.,
\begin{subequations}
\begin{align}
x(t) &= R_1(1-A_1e^{-a_1t}) \cos (\omega_1 t)\\
y(t) &= R_2 (1-A_2e^{-a_2t})\sin (\omega_2 t),
\end{align}
\label{eqno13}
\end{subequations}
\noindent where now different parameters $\{ R_i, A_i,a_i,\omega_i\} $, $i=1,2$ characterize the motion
of the two variables. Because the equations characterizing the system dynamics are not explicitly known in this general case, it is
difficult and in most case impossible to write down coupling schemes in order to investigate
the evolution of two coupled elements. It becomes even more complex in schemes with three or more coupled elements, 
where the phase space trajectory can be as daedalian as a fractal attractor.

\section{Power law approach to limiting orbits in 2D}
\label{sec:powerlaw}

In this section we proceed to the approach of circular and elliptic limiting
orbits when the transient dynamics are of power law type. 
Namely, in Sec.~\ref{sec:power-circle} we consider the case of the circle as a limiting orbit and 
in Sec.~\ref{sec:power-ellipse} we generalize to an elliptic limiting orbit.

\subsection{Power law approach: The circle as a limiting orbit}
\label{sec:power-circle}
In analogy to Eqs.~\ref{eqno03} we may consider the following power law approach to the steady state, 
which for simplicity we consider again as a circular orbit.
\begin{equation}
r(t)=R(1-At^{-b}).
\label{eqno21}
\end{equation}
\noindent In Eq.~\ref{eqno21}, $b>0$ is a positive number which relates to the rate of increase of the orbit with
time. $A$ is a constant as in the previous section and $0<A<1$.
We remind that $A$ cannot take values greater than 1, because this leads to negative values for the radius $r(t)$, which is
unnatural. For $t\to \infty$ the radius of motion tends to $r(t\to \infty)= R = $ const. 
The equations of motion now take the form:
\begin{subequations}
\begin{align}
x(t) &= R(1-At^{-b})  \cos (\omega t)\\
y(t) &= R(1-At^{-b}) \sin (\omega t)\\
x^2(t)& + y^2(t)=r^2(t)=R^2\left( 1-At^{-b}\right) ^2,
\end{align}
\label{eqno22}
\end{subequations}
\noindent where $t\ge t_0>0$. 

\begin{figure}[ht!]
\centering
\includegraphics[width=0.7\linewidth]{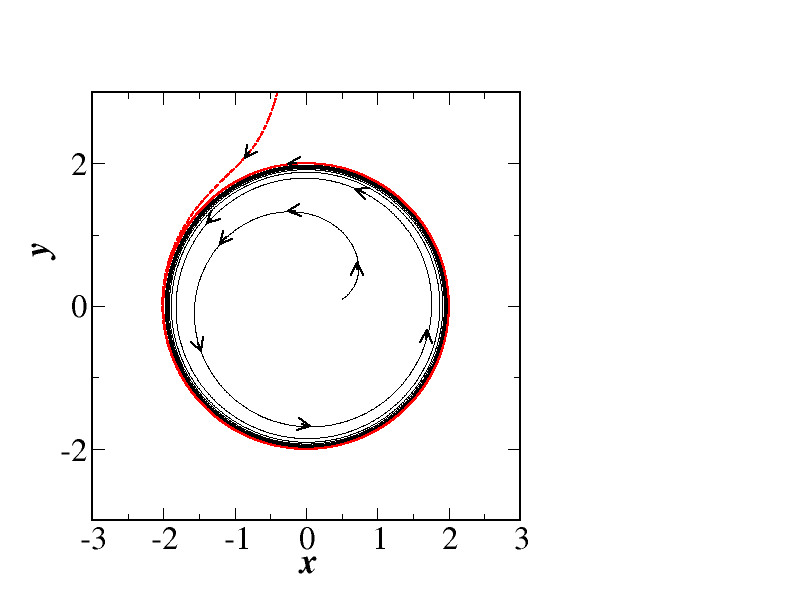}
\caption{(Color online) Phase space diagram representing the variables $x(t)$ and $y(t)$
as they approach their limiting orbit, via a power law (numerical integration of Eqs.~\ref{eqno23}).
Parameters are $b=3.1$, $R=2$ and $\omega =0.3$. The black-solid line and the red-dashed line
correspond to different initial conditions, inside and outside the limiting orbit.}
\label{fig:04}
\end{figure}

The dynamical system whose solutions are the above equations can be written as:
\begin{subequations}
\begin{align}
\frac{dx}{dt} &= \frac{b}{(AR)^{1/b}} \left( R-\sqrt{x^2+y^2}\right) ^{\frac{b+1}{b}}\frac{x}{\sqrt{x^2+y^2}}-\omega y\\
\frac{dy}{dt} &= \frac{b}{(AR)^{1/b}}\left( R-\sqrt{x^2+y^2}\right) ^{\frac{b+1}{b}}\frac{y}{\sqrt{x^2+y^2}}+ \omega x
\end{align}
\label{eqno23}
\end{subequations}
For the derivation of the above equations the following formulas were used:
 $t=\left( (R^2-x^2-y^2)/AR^2\right) ^{-1/b}$, $\cos (\omega t)=x/\sqrt{x^2+y^2}$ and 
$\sin (\omega t)=y/\sqrt{x^2+y^2}$.
Here also, in the case of constant radius, $x^2+y^2=R^2$, the first terms on the right hand side
of Eqs.~\ref{eqno23} vanish and 
consequently the system reduces to Eqs.~\ref{eqno02}, the harmonic oscillator. Figure \ref{fig:04} represents the numerical integration of Eqs.~\ref{eqno23} for two different initial conditions starting outside and inside the final orbit. We note that the trajectory makes many noticeable revolutions before converging to the limiting orbit (at $t\to \infty $). 
\par
When the trajectory starts outside the limiting orbit the radius takes the form $r(t)=R(1+At^{-b})$, as in the case of Sec.~\ref{sec:exp-circle}. To avoid computing fractional powers of negative numbers we reformulate Eqs.~\ref{eqno23} to take explicitly into account the fact that $R^2>(x^2+y^2)$. This reformulation leads to a modified form of Eqs.~\ref{eqno23}, 
which are given in detail in \ref{appendixA}.
\subsection{Power law approach: The ellipse as a limiting orbit}
\label{sec:power-ellipse}
\par
In the same way as in Sec.~\ref{sec:exp-ellipse}, we can derive the equations for the power law approach
to the ellipse. The equations describing the temporal evolution of the variables $\left[ x(t),y(t)\right]$  have the form:
\begin{subequations}
\begin{align}
x(t) &= R_1(1-At^{-b}) \cos (\omega t)\\
y(t) &= R_2 (1-At^{-b})\sin (\omega t)\\
\frac{x^2(t)}{{R_1} ^2}& + \frac{y^2(t)}{{R_2} ^2}=(1-At^{-b})^2.
\end{align}
\label{eqno24}
\end{subequations}

\par Calculations similar to the ones in the previous sections
lead to the following dynamical system:
  
\begin{subequations}
\begin{align}
\frac{dx}{dt} &=
 \frac{b}{A^{1/b}} 
\left( {1-\sqrt{\left( \frac{x}{R_1}\right) ^2+\left( \frac{y}{R_2}\right)^2 }} \right) ^{\frac{b+1}{b}}
\frac{x}{\sqrt{\left( \frac{x}{R_1}\right)^2+\left( \frac{y}{R_2}\right)^2}}-
\frac{R_1}{R_2}\> \omega \> y\\
\frac{dy}{dt} &=  \frac{b}{A^{1/b}}
\left( {1-\sqrt{\left( \frac{x}{R_1}\right) ^2+\left( \frac{y}{R_2}\right)^2 }} \right) ^{\frac{b+1}{b}}\frac{y}{\sqrt{\left( \frac{x}{R_1}\right)^2+\left( \frac{y}{R_2}\right)^2}}+\frac{R_2}{R_1}\> \omega \> x 
\end{align}
\label{eqno25}
\end{subequations}
The above equations are modified slightly for trajectories starting outside the limiting orbit, as is shown in \ref{appendixA} (for the case of the exponential approach).

\section{Comparative Relaxation Properties}
\par In this section we study comparatively the relaxation to the circular limiting orbits with time. 
To show the difference in the two cases we calculate the temporal evolution of the 
relaxation measure $S(t)$ defined as: 
\begin{equation}
S(t)=1-\frac{\sqrt{x^2+y^2}}{R^2}
\label{eqno26}
\end{equation}
From the analytical solutions, Eqs.~\ref{eqno04} and \ref{eqno22} it is clear that in the 
former case
$S(t)$ will follow an exponential decay, while in the latter case power law behavior is expected.
We now verify this behavior using the dynamical systems, Eq.~\ref{eqno05} and \ref{eqno23}, without employing the exact solutions
for $x(t)$ and $y(t)$.

The form of the $S$ measure is appropriate for circular limiting orbits, where $R_1=R_2=R$. In the more general case,
when we want to calculate the type of relaxation (exponential or power law) we use different measures, $Q_1(t)$ or $Q_2(t)$,
calculated as follows: 
\begin{enumerate}
\item Integrate the system for a long time.
\item While integrating determine the time series $t_1, t_2, \cdots$ when the $y-$variable becomes zero. 
\item On this time series, the $\sin$-function becomes 0 and thus the $\cos$-function takes maximum and minimum values, 1 or -1.
\item For the time series $t_1, t_2, \cdots $ calculate the relaxation measures $ Q_1(t_i)= 
| {x(t_i)}| $ and/or $Q_2(t_i)=R-| {x(t_i)}| $ in the case the radius of the limiting orbit is known.
\item In a double-logarithmic scale the power law relaxation will demonstrate a straight line,
while the exponential decay shows straight line in a simple logarithmic scale.
\end{enumerate}

\par In fact, step 2 of the calculation might prove tricky,  because in numerical integration normally we do not 
fall exactly on the zero value of a function (or any other precise value). 
To overcome this difficulty, we determine the times $t_{i+}$ and $t_{i-}$ when the $y-$ variable changes sign.
 We then make the assumption that the approximate time that the $y-$variable passed through 0 is
$t_i= (t_{i+}+t_{i-})/2 $, while at that time the $x-$variable takes the approximate 
value:  $x_i= \left( x(t_{i+})+x(t_{i-})\right) /2$.

\par
\begin{figure}[ht!]
\includegraphics[width=0.45\linewidth]{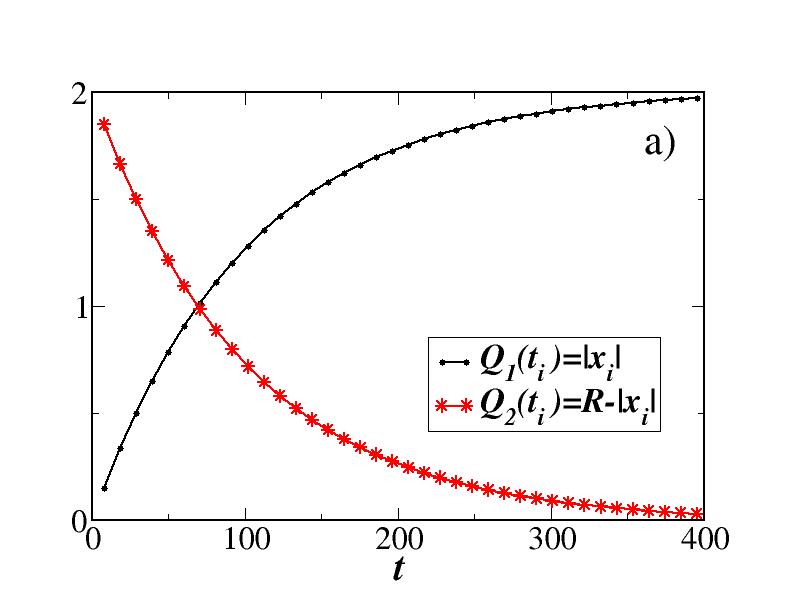}
\includegraphics[width=0.45\linewidth]{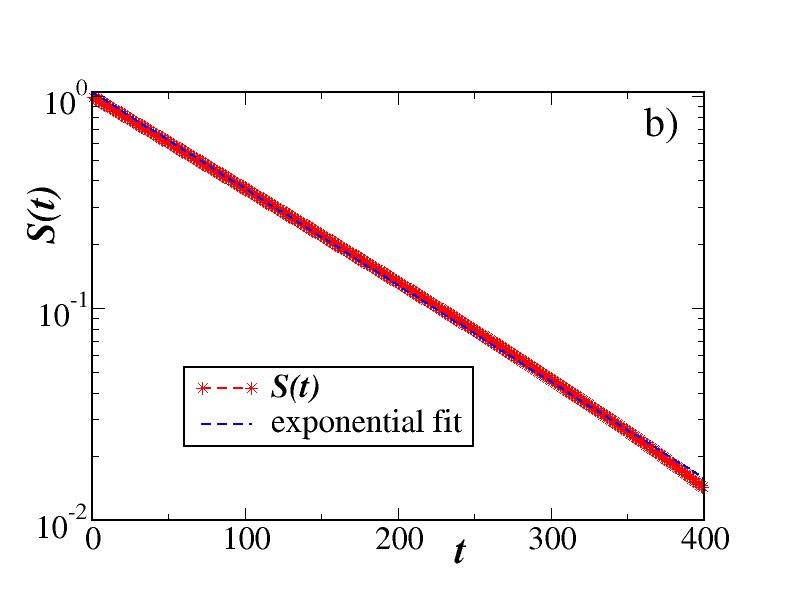}
\caption{(Color online) Temporal evolution of a) the measures $Q(t)$  in normal 
(non-logarithmic) scales and
b) $S(t)$ in single-logarithmic scale. Values from numerical integration of Eqs.~\ref{eqno05}.
Integration parameters are $a=0.01$, $R=2$ and $\omega =0.3$. The  blue-dashed line in b)
designates the exponential fit to the data, with exponent $a_{fit}=0.0105$.}
\label{fig:05}
\end{figure}

 In Fig.~\ref{fig:05} we plot the $Q_{1,2}(t)$ and $S(t)$ measures for system Eqs.~\ref{eqno05},
with parameters $a=0.01$, $R=2$ and $\omega =0.3$. In Fig.~\ref{fig:05}a the $ | {x(t_i)}| $ measure 
approaches the value $R$ for long times, while the $R-|{x(t_i)}|$
measure approaches 0. In a single logarithmic scale, Fig.~\ref{fig:05}b,
 the measure $S(t)$ demonstrates a straight line, 
indicating exponential decay relaxation. The 
exponential fit gives exponent value $a_{fit}=0.0105$, very close to the value $a=0.01$
used in the system integration. While the measures $Q_1(t)$ and $Q_2(t)$ are defined to calculate only deviations 
from one of the two variables ($x$-variable in this case), the measure $S(t)$ takes into account deviations 
of both $x$ and$y$ variables and is equivalent to $Q_2(t)$,
when calculated at time instances $t_i$ because at these times $y(t_i)\sim 0$.
\par
\begin{figure}[ht!]
\includegraphics[width=0.45\linewidth]{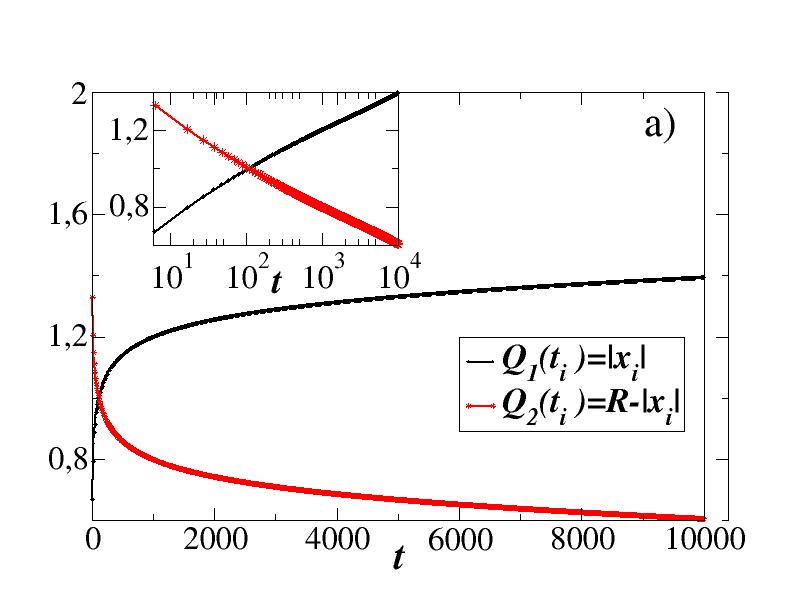}
\includegraphics[width=0.45\linewidth]{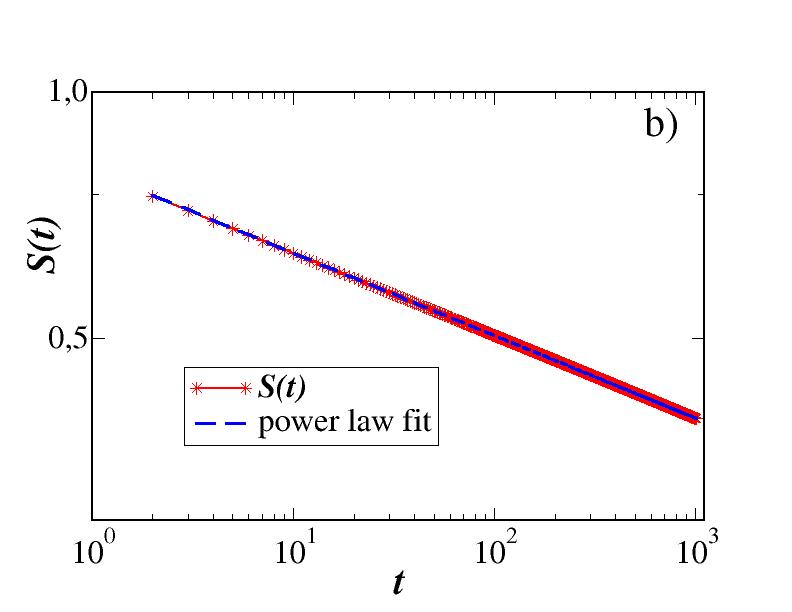}
\caption{(Color online) Temporal evolution of a) the measures $Q(t)$  in normal 
(non-logarithmic) scales and
b) $S(t)$ in double-logarithmic scale (numerical integration of Eqs.~\ref{eqno23}).
The inset in a) shows the linearity of the $Q(t)$ measures in double-logarithmic scale.
 The  blue-dashed line in b)
designates the power law fit to the data, with exponent $b_{fit}=0.1008$.
Parameters are $b=0.1$, $R=2$ and $\omega =0.3$.}
\label{fig:06}
\end{figure}

 Similarly, in Fig.~\ref{fig:06}, the
  $Q_{1,2}(t)$ and $S(t)$ measures for system Eqs.~\ref{eqno23} demonstrate a straight line in double-logarithmic scale, 
indicating power law relaxation. For demonstration purposes we used  $b=0.1$,
10 times higher than in the exponential case, since the power law relaxation is much slower than
the exponential. For the same reason the time axes in Fig.~\ref{fig:06}a,b are in logarithmic scales, 
covering many time scales before the system variables approach their steady states.

\par The constructed measures in single and double-logarithmic scale allow 
to discriminate between the different relaxation approaches to the limiting orbits. 
The measures and the numerical integrations were performed here for the case of 
the circular limiting orbits; similar measures can be constructed for the case of 
approach to the elliptic orbits.

\section{Conclusions and Open Problems}

Dynamical systems were constructed with exactly solvable dynamics 
which demonstrate exponential decay or power law decay to circular or elliptic limiting orbits. 
Adjustable parameters are the
decay exponents and the size of the limiting orbits.

\par The proposed dynamics find applications in natural systems which demonstrate
 time dependent approach to oscillatory steady states. For example, 
in neuron dynamics the single elements demonstrate spiking oscillations 
with abrupt recovery to the rest state. Such oscillations can be achieved
 by assigning a large difference in the two axes of the elliptic orbit. 
Other applications which require power law approach to limiting orbits include
biological cycles, ecological dynamics, while chemical dynamics demonstrate
 mostly exponential decays to the steady state \cite{nicolis:1977,murray:2002}.

\par Another point of interest is whether power law or exponential 
relaxation at the level of the single oscillators 
influences the overall behavior in coupled networks of oscillators.  In particular,
it would be
interesting to study if the type of temporal approach to the steady state modifies
the stability or the morphology or the motion of the chimera state, 
as was discussed in Sec.~\ref{sec:exp-ellipse}.
 It is in our future plans to addresses these questions not only for chimera states but for  
synchronization phenomena in general.

\begin{acknowledgments}
 Computational time for this study was granted from the Greek Research \& Technology Network (GRNET)
in the National HPC facility - ARIS - under project CoBrain3, ID PR005014.
\end{acknowledgments}

\bigskip

\appendix
\section{Initial conditions outside the limiting orbit}
\label{appendixA}
Here we present in detail the form that the Eqs.~\ref{eqno23} take when the trajectory starts 
outside the limiting orbit area. In this case the radius reads:
\begin{equation}
r(t)=R(1+At^{-b}),
\label{eqno61}
\end{equation}
\noindent where $b>0$ is a positive power exponent and $A$ can, now, take values $A>0$.
 For $t\to \infty$ the radius of motion tends  to $R(t\to \infty)=R=$ const. The equations of motion now take the form:
\begin{subequations}
\begin{align}
x(t) &= R(1+At^{-b})  \cos (\omega t)\\
y(t) &= R(1+At^{-b}) \sin (\omega t)\\
x^2(t)& + y^2(t)=r^2(t).
\end{align}
\label{eqno62}
\end{subequations}
\noindent where $t\ge t_0>0$. 
The dynamical system then becomes:
\begin{subequations}
\begin{align}
\frac{dx}{dt} &= - \frac{b}{(AR)^{1/b}} \left( \sqrt{x^2+y^2} - R\right) ^{\frac{b+1}{b}}\frac{x}{\sqrt{x^2+y^2}}-\omega y\\
\frac{dy}{dt} &= -\frac{b}{(AR)^{1/b}}\left( \sqrt{x^2+y^2}-R \right) ^{\frac{b+1}{b}}\frac{y}{\sqrt{x^2+y^2}}+ \omega x
\end{align}
\label{eqno63}
\end{subequations}
This equation has been used of the computations of the outer orbit in Fig.~\ref{fig:04}. Similar modifications apply to Eqs.~\ref{eqno05} and \ref{eqno12} for the calculations of the outer orbits in the corresponding Figs.~\ref{fig:01} and \ref{fig:02}.

\end{document}